\documentclass[prl,aps,tightenlines,twocolumn,superscriptaddress,showpacs,floatfix]{revtex4}
\usepackage{colordvi}
\usepackage{graphicx}
\usepackage{amssymb}
\usepackage{amsmath}
\newcommand{\bgreek}[1]{\mbox{\boldmath$#1$\unboldmath}}
\def\up{\uparrow}
\def\down{\downarrow}

\begin{document}
\title{A virtual intersubband spin-flip spin-orbit coupling induced spin relaxation
in GaAs (110) quantum wells} 
\author{Y.\ Zhou}
\affiliation{Hefei National Laboratory for Physical Sciences at
  Microscale, University of Science and Technology of China, Hefei,
  Anhui, 230026, China} 
\author{M.\ W.\ Wu}
\thanks{Author to whom correspondence should be addressed}
\email{mwwu@ustc.edu.cn.}
\affiliation{Hefei National Laboratory for Physical Sciences at
  Microscale, University of Science and Technology of China, Hefei,
  Anhui, 230026, China} 
\affiliation{Department of Physics,
  University of Science and Technology of China, Hefei,
  Anhui, 230026, China}

\date{\today}

\begin{abstract}
A spin relaxation mechanism is proposed based on a second-order
spin-flip intersubband spin-orbit coupling together with the
spin-conserving scattering. The corresponding spin relaxation time is 
calculated via the Fermi golden rule. It is shown that this
mechanism is important in symmetric GaAs (110) quantum wells with high 
impurity density. The dependences of the spin relaxation time on
electron density, temperature and well width are  studied with
the underlying physics analyzed.
\end{abstract}



\maketitle

In recent years, semiconductor spintronics has become a focus of
intense experimental and theoretical
research~\cite{opt-or,Wolf,spintronics1,spintronics2,spintronics3,spintronics4}.    
One of the key factors for the design of the spin-based device
is to understand the spin relaxation such that the information is well
preserved before required operations are completed. 
In $n$-type zinc-blende semiconductors, like GaAs, the leading spin
relaxation mechanism in most situation is the
D'yakonov-Perel' (DP) mechanism~\cite{Dyakonov1,Dyakonov2,Dyakonov3,Dyakonov4}, 
which is from the joint effects of the momentum scattering and the
momentum-dependent effective magnetic field (inhomogenous
broadening~\cite{Wu_early1,Wu_early2,Weng_03}) induced by the
Dresselhaus~\cite{Dresselhaus_55} and/or
Rashba~\cite{Rashba_84,Rashba_84_2} spin-orbit coupling (SOC).   
However, in symmetric (110)-oriented GaAs quantum wells (QWs), when
only the lowest subband is occupied, the in-plane components of the
spin-orbit field vanish and the effective magnetic field
only exists along the growth direction~\cite{Winkler_04}.
As a result, electrons with spin polarization along the growth direction can
not precess around the effective field, and therefore the DP
mechanism is absent. 
It is also noted that in the presence of an in-plane external magnetic
field, the in-plane and out-of-plane spin relaxations are mixed, thus
the DP mechanism still leads to the spin relaxation/dephasing in this
system, as point out by Wu and Kuwata-Gonokami~\cite{Wu_110}. 
Experimentally Ohno {\em et al.}~\cite{Ohno_99} first observed very
long spin relaxation time (SRT) in GaAs (110) QWs, which exceeds the
SRT in (100) QWs by more than one order of magnitude. Later the
spin dynamics in this system was studied by many works~\cite{Ohno_2000,
Ohno_2001,Dohrmann_04,Hagele_05,gate1,gate2,sysmetry1,sysmetry2,Muller_08}. 
In most of these works, the main reason limiting the SRT is thought to
be the Bir-Aronov-Pikus (BAP) mechanism~\cite{Bir1,Bir2}, which is 
due to the electron-hole exchange interaction 
(holes are from the optical excitation in $n$-type samples).
In the spin noise spectroscopy measurements by M\"uller
{\em et al.}~\cite{Muller_08}, the excitation of semiconductor
is negligible, and hence the BAP mechanism is avoided. 
They reported even longer SRT about $24$~ns and attributed it to
the DP mechanism due to the random Rashba fields caused by
fluctuations in the donor density~\cite{Sherman_APL,Sherman_PRB}. 
In this note, we propose another spin relaxation mechanism to
understand the slow spin relaxation in the {\em absence} of the DP and BAP
mechanisms, which is based on a second-order spin-flip process of the
intersubband SOC together with the spin-conserving scattering. We will show that
this mechanism is important in symmetric  GaAs (110) QWs with high impurity
density.

We start our investigation from symmetrically doped GaAs
(110) QWs with the growth direction along the $z$-axis.
The well width is assumed to be small enough so that only the lowest
subband is occupied for the temperature and electron
density we discuss. 
The envelope functions of the relevant subband are calculated under
the finite-well-depth assumption.
The barrier layer is chosen to be Al$_{0.39}$Ga$_{0.61}$As where the
barrier height is 328~meV~\cite{Yu_92}.
The Hamiltonian can be written as
($\hbar\equiv 1$)
\begin{eqnarray}
  \nonumber
  H\hspace{-0.05cm}&=&\hspace{-0.25cm}
  \sum_{nn^\prime{\bf k}{\sigma}\sigma^{\prime}}\hspace{-0.1cm}\left[
    (\epsilon_{{\bf k}}+E_n)
    \delta_{nn^\prime}\delta_{\sigma\sigma^{\prime}}
    +{\bf h}^{nn^\prime}({\bf k})\cdot\frac{{\bgreek
      \sigma}_{\sigma\sigma^{\prime}}}{2} \right]  \\ &&  
  \times c^{\dagger}_{n{\bf k}\sigma}c_{n^\prime{\bf k}\sigma^{\prime}}
  +H_I.
\end{eqnarray}
Here $\epsilon_{{\bf k}}=k^2/2m^\ast$ is the energy spectrum of the 
electron with two-dimensional momentum ${\bf k}=(k_x,k_y)$;
$E_n$ represents the quantized energy of the electron in the
$n$-th subband; ${\bgreek \sigma}$ is the Pauli matrix. 
In symmetric QWs without external gate voltage, the Rashba SOC can be
neglected~\cite{Inter-Rashba}, and the spin-obit field ${\bf
  h}({\bf k})$ is only from the Dresselhaus
term. In the (110) coordinate system, ${\bf h}({\bf k})$ can be
written as
\begin{eqnarray}
  \nonumber
  h^{nn^\prime}_x({\bf k})&=&\gamma_{\rm D}\left[
  -(k_x^2+2k_y^2)\langle n| k_z|n^\prime\rangle
    + \langle n|k_z^3|n^\prime\rangle \right],\\
  \nonumber
  h^{nn^\prime}_y({\bf k})&=& 4\gamma_{\rm D}k_xk_y \langle n|
  k_z|n^\prime\rangle,\\ 
  h^{nn^\prime}_z({\bf k})&=& \gamma_{\rm D}k_x\left(k_x^2-2k_y^2-
    \langle n|k_z^2|n^\prime\rangle\right) \delta_{nn^\prime},
\end{eqnarray}
where $\gamma_{\rm D}= 30$~eV$\cdot${\AA}$^3$ denotes the Dresselhaus SOC
coefficient~\cite{Rossler} and $\langle n|k_z^{m}|n^\prime\rangle=
\int\,dz\phi_{n}^\ast(z)(-i{\partial}/{\partial}_z)^m\phi_{n^\prime}(z)$
with $\phi_{n}(z)$ representing the envelope function of the
electron in $n$-th subbands. 
Since $\langle n|k_z|n\rangle=\langle n|k_z^3|n\rangle=0$,
all the in-plane components of the intersubband spin-orbit field
vanish as mentioned above. However, $\langle
1|k_z|2\rangle\ne0$, thus the in-plane components of the intersubband
spin-orbit field are still present. The interaction Hamiltonian $H_I$
is composed of the electron-ionized-impurity,
electron-longitudinal-optical (LO)-phonon, electron-acoustic
(AC)-phonon and electron-electron 
Coulomb interactions. Their expressions can be found in
textbooks~\cite{Mahan_1981,Haug_1998}. 

\begin{figure}[htbp]
  \begin{center}
    \includegraphics[width=7cm]{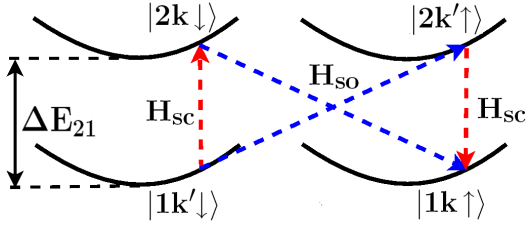}
  \end{center}
  \caption{(Color online) Schematic representation of a second-order
    spin-flip process based on the intersubband spin-orbit field
    $H_{\rm so}$ and the spin-conserving scattering $H_{\rm sc}$. 
  }
  \label{Schematics}
\end{figure}

A second-order spin-flip process can be constructed via the
intersubband spin-orbit field and the spin-conserving scattering as
depicted in Fig.~1. The matrix element of the spin-flip
transition from the electron-impurity scattering (which is referred
to as the spin-flip electron-impurity scattering in the
following) is given by
\begin{eqnarray}
  \nonumber
  U_{\up\down}({\bf k},{\bf k}^\prime,q_z)
  &=&  \frac{\langle 1{\bf k}\up|H_{\rm so}|2{\bf k}
    \down\rangle\hspace{0.1cm} 
    \langle 2{\bf k}\down|H_{\rm ei}^{\rm sc}|1{\bf k}^\prime \down\rangle}
  {{\epsilon}_{{\bf k}^\prime} - {\epsilon}_{{\bf k}}-\Delta E_{21}}
  \\ && \hspace{-0.8cm}  
  {}+ \frac{\langle 1{\bf k}\up|H_{\rm ei}^{\rm sc}
    |2{\bf k}^\prime \up\rangle \hspace{0.1cm} \langle 2{\bf k}^\prime
    \up|H_{\rm so} |1{\bf k}^\prime\down\rangle} 
  {-\Delta E_{21}}.
\end{eqnarray} 
Here $\Delta E_{21}=E_2-E_1$ stands for the energy splitting between
the first and second subbands of electrons. 
Although the above process is similar to the short-range Elliott-Yefet spin
relaxation caused by virtual scattering~\cite{opt-or,Ivchenko},
the intermediate virtual states are chosen to be the states in the high
conduction subband in stead of the  states in the valence band in the
previous work~\cite{Ivchenko}. 
From the energy conservation $\epsilon_{\bf k}=\epsilon_{{\bf k}^\prime}$, the
symmetry of the form factor $I_{12}(q_z)=I_{21}(q_z)=\langle 1|e^{iq_z
z}|2\rangle$ and the symmetry of the spin-orbit field 
${\bf h}^{21}_{\|}({\bf k})=-{\bf h}^{12}_{\|}({\bf k})$, 
$U_{\up\down}({\bf k},{\bf k}^\prime,q_z)$ can be rewritten as
\begin{equation}
  U_{\up\down}({\bf k},{\bf k}^\prime,q_z) =
  \frac{U_{{\bf k}^\prime-{\bf k}}I_{12}(q_z)} 
  {-\Delta E_{21}} [{\bf h}^{12}({\bf k})-{\bf h}^{12}({\bf k}^\prime)]
  \cdot{\bgreek \sigma}_{\up\down}.
  \label{martix_eisf}
\end{equation}
Similarly, one can obtain the matrix element of the spin-flip
electron-phonon scattering
\begin{equation}
D_{\up\down}^\eta({\bf k},{\bf k}^\prime,q_z)=
  \frac{D_{{\bf Q}\eta} I_{12}(q_z)}
  {-\Delta E_{21}} [{\bf h}^{12}({\bf k})-{\bf h}^{12}({\bf k}^\prime)]
  \cdot{\bgreek \sigma}_{\up\down},
  \label{martix_epsf}
\end{equation} 
where ${\bf Q}=({\bf k}^\prime-{\bf k},q_z)$ is the three-dimensional
momentum. $U_{{\bf k}^\prime-{\bf k}}$ and $D_{{\bf Q}\eta}$
represent the matrix elements of the electron-impurity and
electron-phonon interactions respectively, whose expressions are given
in detail in Ref.~\onlinecite{Zhou_PRB_08}. From Eqs.~(\ref{martix_eisf}) and 
(\ref{martix_epsf}), it is seen that the matrix elements of the spin-flip
electron-impurity and electron-phonon scatterings are  
proportional to $|{\bf h}^{12}({\bf k})-{\bf h}^{12}({\bf
  k}^\prime)|/\Delta E_{21}$. Thus the term containing $k_z^3$ in
${\bf h}({\bf k})$, which is independent of two-dimensional momentum ${\bf
  k}$, has no contribution to the spin-flip scattering.

By using the Fermi golden rule, the SRT is given
by ${T_1}^{-1}={T_1^{\rm ei}}^{-1}+{T_1^{\rm ep}}^{-1}$ with~\cite{noeesf}
\begin{equation}
{T_1^{\rm ei/ep}}^{-1}=\frac{2\sum\limits_{{\bf k}{\bf k}^\prime q_z}\Gamma_{\up
    \down}^{\rm ei/ep}({\bf k},{\bf k}^\prime,q_z)f_{{\bf k}}(1-f_{{\bf k}^\prime}) 
}{\sum\limits_{{\bf k}{\bf k}^\prime}f_{{\bf k}}(1-f_{{\bf k}^\prime})}.
\label{T1}
\end{equation}
Here $f_{{\bf k}}$ is the equilibrium electron distribution function.  
$\Gamma_{\up\down}^{\rm ei/ep}({\bf k},{\bf k}^\prime,q_z)$ represents
the transition rate of the spin-flip electron-impurity or
electron-phonon scattering, which can be written as
\begin{eqnarray}
  \Gamma_{\up\down}^{\rm ei}({\bf k},{\bf k}^\prime,q_z)&=& 2\pi 
  |U_{\up\down}({\bf k},{\bf k}^\prime,q_z)|^2\delta({\epsilon}_
  {\bf k} -{\epsilon}_{{\bf k}^\prime}),\\ 
  \nonumber
  \Gamma_{\up\down}^{\rm ep}({\bf k},{\bf k}^\prime,q_z) &=& 2\pi 
  \sum_{\eta} |D_{\up\down}^{\eta}({\bf k},{\bf k}^\prime,q_z)|^2
  \left[N_{{\bf Q}\eta} \delta({\epsilon}_{\bf k}
    -{\epsilon}_{{\bf k}^\prime}  \right. 
  \\ && \hspace{-0.8cm} \left.
    {} +\omega_{{\bf Q}\eta}) 
     + (N_{{\bf Q}\eta}+1)\delta({\epsilon}_{\bf k}
    -{\epsilon}_{{\bf k}^\prime}-\omega_{{\bf Q}\eta})
  \right].
  \label{Gamma_sf}
\end{eqnarray}
Here $N_{{\bf Q}\eta}$ and $\omega_{{\bf Q}\eta}$ are the
distribution function and energy spectrum of the phonon with mode  
$\eta$ and momentum ${\bf Q}$. 
It is noted that the spin relaxation process we discuss is only
from the virtual intersubband scattering. The 
spin relaxation due to the real intersubband
scattering~\cite{Weng_04,Dohrmann_04,Hagele_05} is negligible 
due to the small well width in the present investigation.


\begin{figure}[htbp]
  \begin{center}
    \includegraphics[width=6.5cm]{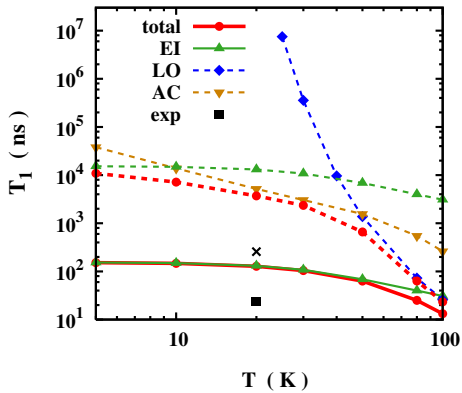}
  \end{center}
  \caption{(Color online) SRTs due to the spin-flip electron-impurity,
    electron-AC-phonon and electron-LO-phonon scatterings  
    as well as the total SRT with all the spin-flip scatterings included
    {\em vs.} temperature $T$ with well width $a=16.8$~nm and electron
    density $N_e=1.8\times10^{11}$~cm$^{-2}$. The impurity densities are
    $N_i=N_e$ (solid curves) and $0.01~N_e$ (dashed curves), respectively.
    The square represents the experimental result in
    Ref.~\onlinecite{Muller_08}. The cross represents the SRT
    due to the RRDP mechanism for $N_i=N_e$, calculated via the
    relation $T_1^{\rm RRDP}\propto 1/\tau_p$. 
  }
  \label{fig_scat}
\end{figure}

\begin{figure}[htbp]
  \begin{center}
    \includegraphics[width=6.5cm]{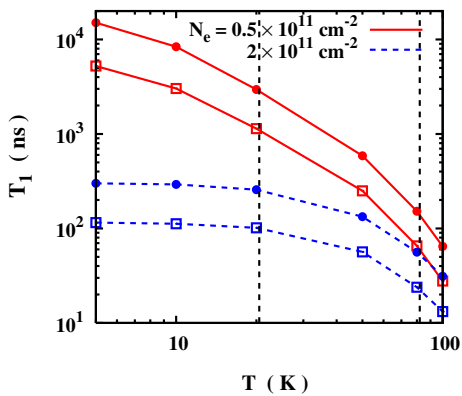}
  \end{center}
  \caption{(Color online) SRTs with all the spin-flip scatterings included
    {\em vs.} temperature $T$ for $N_i=N_e=0.5 \times
    10^{11}$~cm$^{-2}$ and $2 \times 10^{11}$~cm$^{-2}$, respectively.
    The well widths are $a=10$~nm ($\bullet$) and $16.8$~nm
    ($\square$). The black dashed vertical lines indicate the Fermi
    temperatures $T_{\rm F}^e$ for both electron densities.
  }
  \label{fig_T}
\end{figure}

\begin{figure}[htbp]
  \begin{center}
    \includegraphics[width=6.5cm]{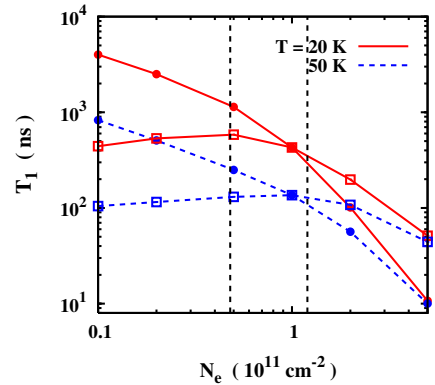}
  \end{center}
  \caption{(Color online) SRTs with all the spin-flip scatterings
    included {\em vs.} electron density $N_e$ for temperatures
    $T=20$~K and $50$~K, with impurity densities $N_i=N_e$ ($\bullet$)
    and $N_i=10^{11}$~cm$^{-2}$ ($\square$). The black dashed vertical
    lines indicate the electron densities satisfying 
    $E_{\rm F}^e=k_{\rm B}T$ for both temperatures. 
  }
  \label{fig_Ne}
\end{figure}

Our main results are summarized in Figs.~2--4.
In Fig.~2, we plot the SRTs due to the spin-flip
electron-impurity, electron-AC-phonon and electron-LO-phonon
scatterings together with the total SRT with all the spin-flip
scatterings included as function of temperature for $N_i=N_e$ (solid
curves) and $N_i=0.01~N_e$ (dashed curves). $N_e=1.8\times
10^{11}$~cm$^{-2}$ and $a=16.8$~nm. In the low impurity density case
with $N_i=0.01~N_e$, it is seen that the spin relaxation due to the
spin-flip electron-AC-phonon and
spin-flip electron-impurity scatterings are comparable at low
temperature and the contribution from the spin-flip electron-LO-phonon 
scattering becomes more important and even dominates at high
temperature. In the high impurity density case with $N_i=N_e$, the spin relaxation
due to the spin-flip electron-impurity scattering is dominant
at most temperatures we 
discuss, and the contribution from the spin-flip electron-LO-phonon
scattering becomes comparable to that from the spin-flip 
electron-impurity scattering when $T>80$~K.

We also compare our results with the experiment by M\"uller {\em et
  al.}~\cite{Muller_08} (square) in Fig.~2. By
fitting the measured mobility given in 
Ref.~\onlinecite{Muller_08}, one obtains $N_i\sim 0.01~N_e$. 
In this case, it is seen that the SRT obtained from our model is two
orders of magnitude larger than the experimental data at $T=20$~K.
However, in the high impurity 
case with $N_i=N_e$, it is shown that the spin relaxation rate due to
our mechanism is enhanced by more than one order of magnitude. 
Meanwhile, the spin relaxation rate due to the random Rashba field
induced DP (RRDP) mechanism~\cite{Sherman_APL,Sherman_PRB} is
suppressed significantly due to the 
increase of the electron-impurity scattering. 
Here we estimate the SRT due to the RRDP
mechanism at $N_i=N_e$ (cross) by using the relation
$T_1^{\rm RRDP}\propto 1/\tau_p$~\cite{opt-or}.
It is seen that the mechanism from our model becomes more important
compared with the RRDP mechanism in the high impurity case. 

Now we discuss the temperature dependence of the SRT. 
In Fig.~3 we plot the SRT as 
function of temperature for different electron densities and well
widths. It is shown that the SRT first decreases slowly and then
rapidly with $T$. The turning point is roughly around the electron Fermi
temperature $T_{\rm F}^{e}=E_{\rm F}^{e}/k_{\rm B}$. 
Since the electron-impurity scattering has a weak temperature
dependence, the temperature dependence of the SRT is mainly from the
$k$-dependent spin-orbit field ${\bf h}({\bf k})$. From
Eqs.~(\ref{martix_eisf}) and (\ref{martix_epsf}), it is seen that only
the quadratic terms in ${\bf h}({\bf k})$ contribute to the spin-flip scattering.
When temperature increases, more electrons are distributed at lager
momentum states, and then the contribution from ${\bf h}({\bf k})$ becomes 
larger. This leads to a larger spin-flip scattering rate and hence
reduces the SRT. It is also noted the average momentum of electrons
is not sensitive to temperature in the degenerate regime 
($T<T_{\rm F}^{e}$). Thus the SRT decreases slowly at low temperature.
From Fig.~3, it is also seen that the SRT becomes shorter
for wider well width. It can be understood as follows. Since the form
factor is weakly dependent on well width, one obtains that 
$T_1 \propto (\Delta E_{21}/\langle 1|k_z|2\rangle)^2$ 
from Eqs.~(\ref{martix_eisf})--(\ref{Gamma_sf}).
It is known that $\Delta E_{21}\propto a^{-2}$ and $\langle
1|k_z|2\rangle\propto a^{-1} $ under the infinite well-depth
assumption. Thus the SRT decreases with an increase  
of well width. 

The electron density dependence of the SRT is also investigated.
In Fig.~4, the SRT is plotted as function of electron
density for different temperatures and impurity densities. 
We first discuss the case with fixed impurity density
$N_i=10^{11}$~cm$^{-2}$. It is seen that the SRT first increases and
then decreases with $N_e$ with a peak around the electron density 
satisfying $E_{\rm F}^e=k_{\rm B}T$. This peak originates from the
competition of two effects. On one hand, similar to the discussion in
temperature dependence, with the increase of electron density,
the average $k$ increases and thus the contribution from
${\bf h}({\bf k})$ is enhanced. 
This effect leads to a decrease of the SRT.
On the other hand, the screening of electrons also increases with
electron density, which suppresses the spin-flip scattering
and increases the SRT. 
When the electron density is low and $E_{\rm F}^e<k_{\rm  B}T$, 
i.e., in the nondegenerate regime, the average $k$ changes
little with $N_e$. Thus the effect of the increase in the screening is
dominant and the SRT increases with increasing density. 
However, when the electron density is high
enough so that $E_{\rm F}^e>k_{\rm  B}T$, i.e., in the degenerate
regime, the effect of the increase of the contribution from ${\bf h}
({\bf k})$ becomes more important. Consequently the SRT
decreases. Then we turn to the case with $N_i=N_e$. It is seen that
the SRT decreases monotonically with electron density. This is because 
the effect of the increase of the impurity scattering 
surpasses the one of the 
increase of screening, and makes the SRT decrease even in low electron
density (nondegenerate) regime. Nevertheless, one still can see that
the SRT decreases more rapidly in the degenerate regime.

In conclusion, we have proposed a spin relaxation mechanism based on a
second-order spin-flip process of the intersubband spin-flip SOC
and the spin-conserving scattering, and calculated the corresponding
SRT via the Fermi golden rule. We show that this mechanism
is important in symmetric GaAs (110) QWs with high impurity density. 
The temperature, well width and electron density dependences of the SRT
are also investigated.

This work was supported by the National Natural Science Foundation of
China under Grant No.\ 10725417, the National Basic
Research Program of China under Grant No.\ 2006CB922005 and the
Knowledge Innovation Project of Chinese Academy of Sciences.
We thank M. Q. Weng for valuable discussions.

\end{document}